\documentclass[5p]{elsarticle}
\usepackage{epsfig}
\begin{document}

\title{Nuclear stopping and rapidity loss in Au+Au collisions at $\sqrt{s_{NN}}$=62.4 GeV}

\date{\Version}
%\date{\today}

\author[oslo]{I.C.Arsene}
%\author[nbi]{I.G.Bearden}}
\author[nbi]{I.G.Bearden}
\author[bnl]{D.Beavis}
\author[uk]{S.Bekele}
\author[buca]{ C.Besliu}
\author[nyu]{B.Budick}
\author[nbi]{H.B{\o}ggild}
 \author[bnl]{C.Chasman}
 \author[nbi]{C.H.Christensen}
 \author[nbi]{P.Christiansen\fnref{dddag}}
 \author[nbi]{H.H.Dalsgaard\corref{cor1}}
 \author[bnl]{R.Debbe}
\author[nbi]{{J.J.Gaardh{\o}je}}
\author[tamu]{ K.Hagel}
\author[bnl]{ H.Ito}
\author[buca]{ A.Jipa}
 \author[uk]{E.B.Johnson\fnref{dag}}
 \author[nbi]{ C.E.J{\o}rgensen}
 \author[krakow]{ R.Karabowicz}
 \author[krakow]{ N.Katrynska}
 \author[uk]{E.J.Kim\fnref{ddag}}
 \author[nbi]{ T.M.Larsen}
 \author[bnl]{ J.H.Lee}
 \author[oslo]{ G.L{\o}vh{\o}iden}
 \author[krakow]{ Z.Majka}
 \author[uk]{ M.J.Murray}
 \author[tamu]{ J.Natowitz}
 \author[nbi]{ B.S.Nielsen}
 \author[nbi]{ C.Nygaard}
 \author[uk]{ D.Pal}
 \author[oslo]{A.Qviller}
 \author[ichp]{F.Rami}
 \author[nbi]{ C.Ristea}
 \author[buca]{ O.Ristea}
 \author[uib]{ D.R{\"o}hrich}
 \author[uk]{ S.J.Sanders}
 \author[krakow]{ P.Staszel}
 \author[oslo]{ T.S.Tveter}
\author[bnl]{F.Videb{\ae}k\fnref{sp}}
 \author[tamu]{ R.Wada}
 \author[uib]{ H.Yang}
 \author[uib]{Z.Yin\fnref{ddddag}}
\author[ss]{I.S.Zgura}

\address[bnl]{Brookhaven National Laboratory, Upton, New York, USA}
\address[iphc]{Institut Pluridisciplinaire Hubert Curien et Universit{\'e} Louis
  Pasteur, Strasbourg, France}
\address[ss]{Institute of Space Science,Bucharest-Magurele, Romania}
\address[krakow]{M. Smoluchowski Inst. of Physics,  Jagiellonian University, Krakow, Poland}
\address[nyu]{New York University, New  York, USA,}
\address[nbi]{Niels Bohr Institute,  University of Copenhagen, Copenhagen, Denmark}
\address[tamu]{Texas A$\&$M University, College Station, Texas, USA}
\address[uib]{University of Bergen, Department of Physics and Technology, Bergen,  Norway}
\address[buca]{University of Bucharest, Romania}
\address[uk]{University of Kansas, Lawrence, Kansas, USA }
\address[oslo]{University of Oslo, Department of Physics, Oslo, Norway}

\fntext[dag]{Present address: Radiation Monitoring Devices, Cambridge, MA, USA}
\fntext[ddag]{ Present address:  Division of Science Education, Chonbuk National University, Jeonju, Korea }
\fntext[dddag]{Present Address: Div. of Experimental
High-Energy Physics, Lund University, Lund, Sweden}
\fntext[ddddag]{Present Address: Institute of Particle Physics, Huazhong Normal University,Wuhan,China}
\fntext[sp]{Spokesperson \it{e-mail: videbaek@bnl.gov}}
\cortext[cor1]{Corresponding Author.\\ \it{e-mail address: canute@nbi.dk}(H.H.Dalsgaard)}

\begin{abstract}
 Transverse momentum spectra of protons and anti-protons measured in the rapidity range $0<y<3.1$ from 0-10\%
  central Au+Au collisions at $\sqrt{s_{NN}}=62.4$ GeV are presented.
   The rapidity densities, $dN/dy$, of protons, anti-protons and
  net-protons $(N_p-N_{\bar{p}})$ have been deduced from the spectra over a rapidity range wide enough to
  observe the expected maximum net-baryon density. 
From mid-rapidity to $y=1$ the net-proton yield is roughly constant ($dN/dy \sim 10$), 
but rises to $dN/dy \sim  25$ at $2.3<y<3.1$. 
The mean rapidity loss  is $2.01\pm 0.16$ units from beam rapidity.
The measured rapidity distributions are compared to model predictions.
Systematics of net-baryon distributions and rapidity loss vs. collision energy are discussed. 
  
PACS numbers: 25.75 Dw.
\end{abstract}

\maketitle
%\clearpage
%%\twocolumn
In collisions between gold nuclei at the top energy
($\sqrt{s_{NN}}=200$ GeV) of the Relativistic Heavy Ion Collider,
RHIC, there is strong evidence of a state of matter
characterized by partonic (quark and gluon) degrees of freedom and
with properties similar to that of a nearly perfect liquid~\cite{BRAHMSwp,STARwp,PHENIXwp,PHOBOSwp,McLerranGyulassy}. The
partons are produced copiously during the initial stages of the
collisions and subsequently hadronize into the roughly 7000~\cite{BRAHMSmult} 
particles produced in central collisions.
The energy required for producing these particles 
comes from the kinetic energy lost by the baryons in the colliding
nuclei. Since $E=m_T\cosh{y}$, where $m_T=\sqrt{m^2+p_T^2}$ is the
transverse mass ($p_T$ is the transverse momentum), and $y$ is the
rapidity, this energy loss is manifest as a loss in mean rapidity of
these baryons.

The net-baryon yield can be estimated from the net-proton yield, which is taken as the difference
between the measured yields of protons and anti-protons.
The rapidity distribution of the net--protons 
%(or, ideally, the number
%of net-baryons if neutral particles are detected) 
after the
collision then not only determines the energy available
for particle production, but also yields information on the
stopping of the ions due to their mutual interaction. 
%We define the `net--proton' yield to be the difference between the measured yields of protons and anti--protons. From this, we estimate the net--baryon yield. 
%By having measurements at energies lower than 200 GeV
% the development of rapidity and energy loss can be studied in greater detail.

By having a measurement at an energy between the lower energy
AGS ($\sqrt{s_{NN}}=5$ GeV) and SPS ($\sqrt{s_{NN}}=17 $ GeV) data and the highest energy RHIC results
($\sqrt{s_{NN}}=200 $ GeV)~\cite{AGSE917,SPSNA49, BRAHMS}, 
the development of rapidity and energy loss can be studied in greater detail.
In this Letter, we present the first measurements of the rapidity
loss of Au ions after central collisions at $\sqrt{s_{NN}}=62.4$ GeV. 
The experimental
arrangement of the BRAHMS detector at RHIC makes it possible to measure
the distribution of net-protons over a rapidity interval $y=0$ to
$y=3.1$. This rapidity range is wide enough to include the maximum
 proton rapidity density, in contrast to the situation
at the RHIC top energy, where the beam rapidity is higher
($y_{b}=5.4$ compared to 4.2 at 62 GeV) and where the acceptance of existing experiments does
not include this peak. The situation at the lower energy therefore
makes it possible to better determine the rapidity density distribution. 
Together with similar information from experiments at lower energies, 
and at the RHIC top energy, we
conclude that the mean rapidity loss of ultra-relativistic heavy ions
exhibits a slowly varying behavior as a function of beam energy
%(which is plotted as projectile rapidity below following the relation above) 
from SPS energies on upwards.

The BRAHMS detector consists of two magnetic spectrometers: the  Mid
Rapidity Spectrometer (MRS) able to cover polar angles (measured
with respect to the beam direction) $ 30^{\circ}<\theta<90^{\circ}
$ and the Forward Spectrometer (FS) able to cover
$2.3^{\circ}<\theta<15^{\circ}$. 
Each spectrometer determines trajectories and momenta of charged hadrons.  
Two Time Projection Chambers (TPCs) are utilized in the MRS and
two TPCs and three Drift Chambers (DCs) in the FS.
Together, they measure protons and anti-protons in the range $-0.1<y<3.5$. 
Collision centrality is determined using a silicon and plastic tile
multiplicity array located around the nominal intersection point (NIP)~\cite{BRAHMSmult}. For
this analysis, the centrality class $0-10\%$ was selected, corresponding to a calculated 
number of participant nucleons of $N_{part}=314 \pm 8$. The
vertex position is determined with an accuracy of $\approx 1$ cm
using two arrays of Cherenkov counters positioned on either side
of the NIP~\cite{BRAHMSNIM}. We have selected
for analysis tracks in the MRS (FS) with vertices within $\pm
15(20) \textrm{cm}$ from the NIP.

\begin{figure}[ht]
  \epsfig{file=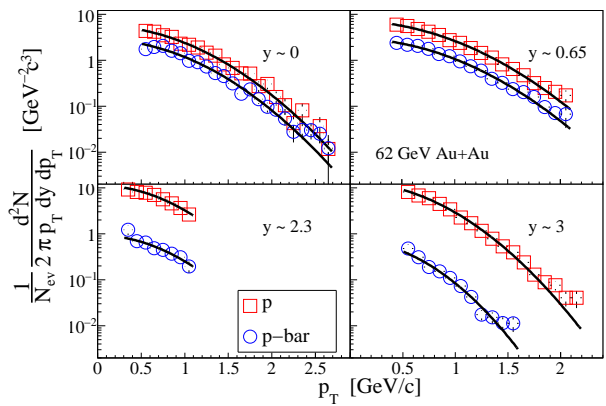,width=\linewidth}
  \caption{Spectra of identified protons and anti-protons for $y \sim
  0$, $y \sim 0.65$, $y \sim 2.3$ and $y \sim 3$ respectively. The solid drawn lines
   are the fit functions used to determine the
  yield. Vertical bars show statistical errors only.}
 \label{fig1}
\end{figure}

Particle identification (PID) is done in the MRS via time of
flight (TOF) measurements, enabling clean identification of protons and anti-protons
in the momentum range $0.4 \textrm{ GeV/}c<p<3 \textrm{ GeV/}c$.
In the FS, two PID detectors are used: A TOF system and a Ring
Imaging Cherenkov (RICH) detector. In the RICH, only (anti-)protons
with $p>15 \textrm{ GeV/}c$ will create resolvable rings.  %of resolvable radius. 
Lower momentum
(anti)protons have no associated Cherenkov radiation. 
In the momentum range $3 \textrm{ GeV/}c < p < 7.5 \textrm{ GeV/}c$ 
we apply a 2 sigma cut about the (anti-) proton peak in the calculated mass spectrum based on the TOF measurement and require that the particle 
not be identified in the RICH as a pion. 
In the
range $12 \textrm{ GeV/}c < p < 20 \textrm{ GeV/}c$ the RICH
is used for PID. For $12 \textrm{ GeV/}c < p < 16 \textrm{ GeV/}c$
the (anti)protons are those particles with either no associated ring or a ring with a small radius.   
Since the RICH is 97\% efficient for particles above threshold, up to 3\% of pions and kaons in this momentum range will be mistakenly identified as protons, an effect for which we correct the data. 

\begin{figure}[ht]
  \epsfig{file=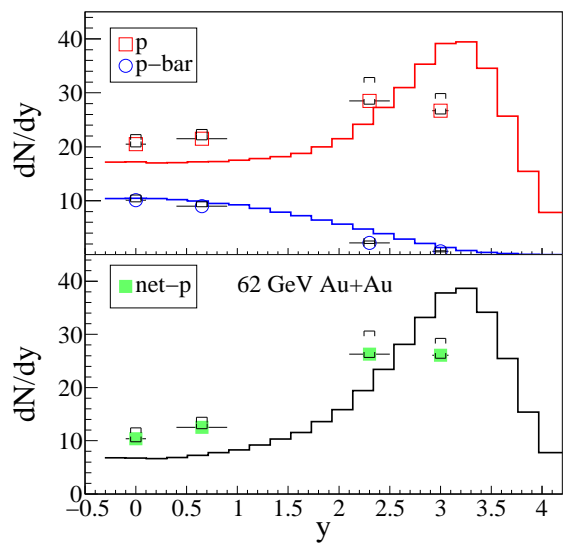,width=\linewidth}
  \caption{Top: rapidity densities of protons and anti-protons. Bottom: rapidity densities of
   net protons. The statistical errors are smaller than the marker
   sizes. The verticalal brackets shows
    the estimated systematic errors due the extrapolation of yields, and the horizontal bars indicate the width of the rapidity interval for each data point. The histograms are from HIJING/$\textrm{B}\bar{\textrm{B}}$(v2.1).}
  \label{fig2}
 \vspace{4mm}
\end{figure}
The differential invariant yields, $\frac{1}{2\pi p_T}\frac{d^2N}{dy dp_T}$, 
of protons and anti-protons have been corrected for
geometrical acceptance and detector efficiency. The acceptance correction is, due to the small solid angle of the BRAHMS spectrometers, the largest correction and is obtained from a purely geometrical simulation 
of each combination of angle and magnetic field used in the experiment. 
The PID efficiency of the TOF
walls is found to be 93-98\%. The total tracking efficiency in
the MRS and FS is $\approx 90 \%$ and $\approx 80 \%$,
respectively. The data have also been corrected for absorption, energy loss and multiple
scattering estimated through simulations using GEANT ~\cite{GEANT}.
The corrections for these physics effects amount to less than 20\% at
the lowest $p_T$ and less than 15\% at the highest $p_T$ for the FS, whereas in the MRS the correction at the highest $p_T$ is $\approx 8\%$.

Fig.~\ref{fig1} shows the invariant spectra of protons and
anti-protons for $y \sim  0$, $y \sim 0.65$, $y \sim 2.3$ and $y
\sim 3$ as a function of  $p_T$. 
The integrated yields have been obtained by fitting these spectra with Gaussian functions ($f(p_T) \propto
exp(-p_T^2/2\sigma^2))$ and integrating the fit functions in the range
$0<p_T<\infty$. From the yields we obtain the rapidity densities
$\frac{dN}{dy}$. The integrated yield under the data points
compared to the extrapolated yield is $\sim 80-90\%$ in the MRS
and $\sim 40-50\%$ in the FS. The mean $p_T$ and $m_T$ determined from the
distributions varies from $\langle p_T \rangle \sim 1 \textrm{ GeV/}c$  
($\langle m_T \rangle \sim 1.4 \textrm{ GeV}$) in the MRS
to $\langle p_T \rangle \sim 0.7 \textrm{ GeV/}c$ ($\langle m_T \rangle \sim 1.2 \textrm{ GeV}$)in the FS.

The top panel of Fig.~\ref{fig2} shows the 
$dN/dy$ of protons and anti-protons. The lower panel shows the
$dN/dy$ of the net-protons, defined here as the difference of the
protons and anti-proton densities. The systematic errors due to the limited $p_T$ coverage have been
determined by varying the fit range, the bin width, the choice of
fit function and the rapidity cuts.  
These uncertainties together with those for the corrections results in an estimate of the total systematic error of roughly 10\%.

The predictions of the 
HIJING/$\textrm{B}\bar{\textrm{B}}$(v2.1) event generator~\cite{HIJING} for the proton, anti-proton and
net-proton rapidity distributions are shown in Fig.~\ref{fig2} by the histograms.
The HIJING distributions include protons and anti-protons from hyperon
decays that would be measured as primary protons (e.g. $0.53 \pm
0.05$ (anti-)protons for each (anti-)lambda, see~\cite{BRAHMS}). 
The HIJING calculation predicts a smaller net-proton yield at mid-rapidity, and shows a peak slightly forward of 
that indicated by our results; thus, the mean rapidity loss is smaller than 
estimated from our data.
%In particular the calculated
%distribution underestimates the net proton yield measured at
%$y=2.3$

The estimate for net-baryons depends on the relative acceptance for
direct and decay protons in the spectrometers and the ratio of
hyperons to protons.  
The acceptances are calculated using GEANT while
the hyperon:proton ratios have been calculated using the
THERMUS~\cite{THERMUS} model.  
The THERMUS results are based on fits
to BRAHMS $\pi^\pm, K^\pm, p$ and ${\bar p}$
data~\cite{Stiles:2006sa}. 
The $n/p$ ratio is obtained from HIJING. 
The necessary conversion from the net-protons, $N_p$, to the net-baryons,
$N_{B}$, is then done (as in Ref.~\cite{BRAHMS}) using $N_{B} = (2 \pm
0.1)\cdot N_p$ at mid-rapidity and $N_{B} = (2.1 \pm 0.1)\cdot N_p$ at
forward rapidities (the larger correction at forward rapidity is due
to a small increase in the $n/p$ ratio at forward rapidities).

\begin{figure}[ht]
  \epsfig{file=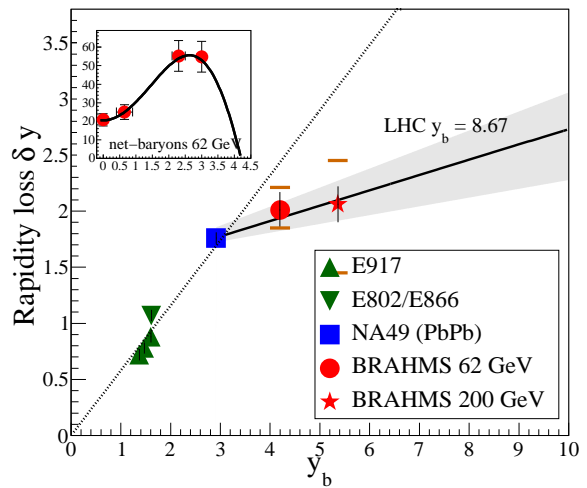,width=1.\linewidth}
  \caption{Rapidity losses from AGS, SPS and RHIC as a function of beam rapidity.
   The solid line is a fit to SPS and RHIC data, and the band is the statistical uncertainty of this fit.  The dashed line is a 
  linear fit to AGS and SPS data from ~\cite{VidebaekHansen}.}
  \label{fig3}
\end{figure}

We quantify  %nuclear stopping using 
the mean rapidity ~\cite{VidebaekHansen} and energy  loss using the following integrals:
\begin{eqnarray}
\delta y & = &  y_{b} -\frac{2}{N_{part}} \int_{0}^{y_{b}} y\frac{dN_{B-\bar{B}}}{dy} dy \\
\delta E & = & E_{b} -\frac{2}{N_{part}} \int_{0}^{y_{b}}
\langle m_T  \rangle \cosh{y} \frac{dN_{B-\bar{B}}}{dy} dy. 
\label{eq:AvRapLoss}
\end{eqnarray}
Here $y_{b}$  and $E_{b}$ are the rapidity and energy of the incoming beams  and $dN_{B-\bar{B}}/dy$ 
is the net-baryon rapidity density. 
Since we only have data at four rapidities, $y=0, 0.65, 2.3$ and 3,  
we have to both interpolate and extrapolate our data to evaluate the integrals. 
We found that $\langle m_T  \rangle $ drops linearly with rapidity. 
For the baryon yield we fitted our data to a   $3^{\rm rd}$ order polynomial in $y^2$ (analogous to ~\cite{BRAHMS}) subject to the constraint that $N_{part}= 314$ and that the yield at $y_b$ is 0. 
The result of this fit together with our net-baryon distribution are shown in the inset of  Fig.~\ref{fig3}.

Using this method we obtain the average rapidity  and
energy loss per participant baryon for Au+Au collisions at $\sqrt{s_{NN}}=62.4
\textrm{ GeV}$:
$$\delta y = 2.01 \pm 0.16 , \quad \delta E = 22 \pm 1 \textrm{ GeV}. $$
Since the energy loss per pair of participant nucleons is $2 \delta E$,
approximately 70\% of the initial beam energy is available for particle production after the collision.

The rapidity losses at AGS ~\cite{AGSE917,AGSE802,AGSE877}, SPS~\cite{SPSNA49} 
and RHIC $\sqrt{s_{NN}}=200 \textrm{ GeV}$
~\cite{BRAHMS} together with the present result are summarized in
Fig.~\ref{fig3}. The figure shows that rapidity loss increases
rapidly with beam energy from AGS to SPS but much more slowly
from SPS energy to RHIC energies. 
The dotted line in Fig.~\ref{fig3} is taken from ~\cite{VidebaekHansen}
where there was found to be a linear scaling over a wide range of
energies up to the SPS top energy. This scaling was found to be broken
at $\sqrt{s_{NN}} = 200$ GeV in ~\cite{BRAHMS} and we conclude that the proposed linear scaling breaks between $\sqrt{s_{NN}} = 17$ GeV and  $\sqrt{s_{NN}} = 62.4$ GeV.      
The solid drawn line is a linear fit to SPS and RHIC data and allows one to extrapolate to 
the LHC regime ($y_{b}=8.7$). The grey band gives the statistical uncertainty of this extrapolation, but is only useful to the extent that the underlying physics is the same from RHIC to LHC energies. 

\begin{figure}[ht]
  \epsfig{file=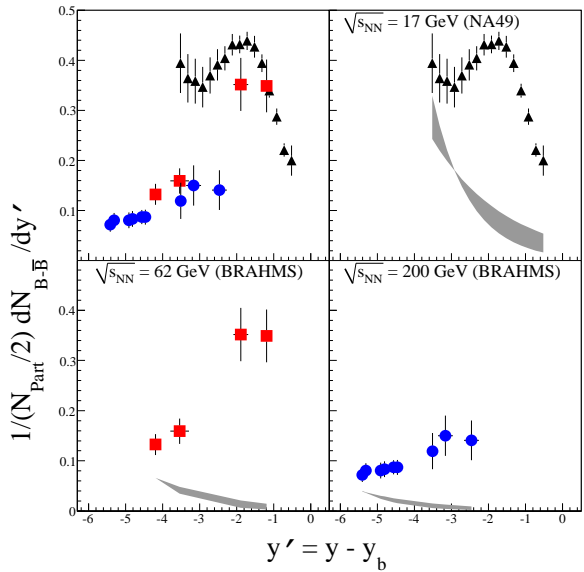,width=1.\linewidth}
  \caption{The top left panel shows $(1/N_{part}/2)dN_{B-\bar{B}}/dy'$ where $y'=y-y_{b}$ for
  SPS and RHIC energies. No scaling is observed. The three other
  panels show data from SPS and RHIC drawn together with the `target'
  net-baryon contribution from ~\cite{Kopeliovich:1988qm}. 
The triangles represent data from NA49, the squares BRAHMS data from 62 GeV and the circles BRAHMS 200 GeV data.}
  \label{fig4}
\end{figure}

Applying the conversion from net-protons to net-baryons and fitting
the result, we estimate that for collisions with $N_{part}\sim314$,
there are $\sim 240$ net--baryons in the acceptance covered by BRAHMS,
$-3.1 < y <  3.1$ (because of symmetry). To estimate the minimum
and maximum possible stopping at $\sqrt{s_{NN}} = 62.4$ GeV the
remaining baryons can be placed at $y = 2.7$ or $y = y_{b}$.  These
values are indicated by the horizontal bars in Fig.~\ref{fig3}. The
horizontal bars for the $\sqrt{s_{NN}} = 200$ GeV are obtained in the
same way, and similarly give the absolute limits on the amount of
stopping.  
If instead of relying on these limits, we use the
uncertainty on the fits to the yields to
perform the extrapolation to LHC, we then find that the rapidity loss at LHC  would be expected to lie between $2.1<\delta y<2.4$.

The slow increase of the rapidity loss from the top SPS energy to the top
RHIC energy indicates that baryon transport does not depend strongly
on energy at high energies when observed in the rapidity frame of the
beam, $y' = y_{CM} - y_{b}$. We therefore compare
$dN/dy'$ for net-baryons normalized to the number of participant pairs. For NA49 data~\cite{SPSNA49} the number of net-baryons is $ 352
\pm 12$ for the interval $|y_{cm}| <2.5$. Extrapolating to $y_{b} =
2.9$ we estimate the number of participants to be $390 \pm 20$.

Fig.~\ref{fig4} (top left) shows net-brayon rapidity densities $1/(N_{part}/2)dN_{B-\bar{B}}/dy'$ for different beam energies. 
The $\sqrt{s_{NN}}=17$ GeV and $\sqrt{s_{NN}}=62.4$ GeV data differ significantly at $y' \sim -3$ but
approach each other towards $y_{b}$, coinciding at
$y' \sim -1$, reflecting the much larger contribution of `target' baryons at $\sqrt{s_{NN}}=17$ GeV. This is  reminiscent of limiting fragmentation and indeed is the reason for choosing the variable $y'$.

\begin{figure}[ht]
  \epsfig{file=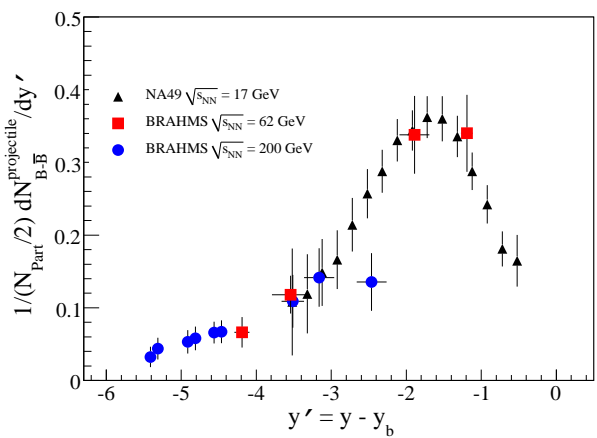,width=1.\linewidth}
  \caption{Projectile net-baryon rapidity density $(1/N_{part}/2)dN^{projectile}_{B-\bar{B}}/dy'$ from SPS and RHIC
    after subtraction of the target net-baryon contribution (see Fig.~\ref{fig4}).}
  \label{fig5}
\end{figure}

This direct comparison of data is complicated by the aforementioned target contribution.
For $\sqrt{s_{NN}}=17$ GeV  at $y'=3$ the target contribution  is half the
net-baryons while for $\sqrt{s_{NN}}=62.4$ GeV and $\sqrt{s_{NN}}=200$
GeV the contribution is significantly
less. To compare the net-baryons from the projectile only, the target contribution 
must be subtracted. 
At $y_{CM} = 0$ the target nucleons comprise half of the net-baryons due to beam-target
symmetry, and at $y_{b}$ the target contribution is expected to be negligible.
Between these  extremes we consider two
different rapidity dependences for the correction: (1) a simple exponential form $\exp(-y')$~\cite{BuszaGoldhaber84} and (2) a gluon
junction motivated form $\exp(-y'/2)$~\cite{Kopeliovich:1988qm}.  
These two functions give limits for the target net-baryon tail 
and provide the bounds for the grey bands in Fig.~\ref{fig4}.  
We subtract the average of these two functions, with half the difference between them taken to be the systematic uncertainty on the function, from the data.
A consequence of the forward-backward (beam-projectile) symmetry is that there is no error on the beam contribution at mid--rapidity.
As one approaches the projectile rapidity, the uncertainty grows, but both the absolute and relative contribution of the target net-baryons decreases and the correction becomes smaller. 
After subtracting the target net-baryon contribution shown in
Fig.~\ref{fig4}, we obtain the  projectile net-baryon distributions
presented in Fig.~\ref{fig5}.  
One notes a remarkable similarity
among the data sets for large rapidity losses $y'<-3$, while the 200 GeV data may begin to diverge from the lower energy data for $y'>-3$.
This suggests that the rapidity loss obtained from Eq.~1, which does not distinguish the target and projectile contributions to  
the net-baryon yield, while experimentally practical, is slightly misleading, 
and that the increase in projectile rapidity loss from SPS to RHIC maximum energy might be smaller than what is implied in Fig.~\ref{fig3}.

Finally, we note that the present results for the rapidity loss of
baryons(protons) for Au+Au collisions at $\sqrt{s_{NN}}=62.4$ GeV,
together with similar data at lower and higher energy permit us to
more accurately determine the onset of a collision regime where
the absolute rapidity loss appears to vary slowly with beam
rapidity, with the data exhibiting  a near saturation of the rapidity loss
with increasing collision energy. 
Recently, it has been argued~\cite{Mehtar-Tani08} that such a scaling may result from gluon saturation of the collision.   
These results suggest  that
the rapidity loss at LHC will be approximately $\delta y = 2.3$.
Although the baryon rapidity loss relative to the beam rapidity
decreases, the overall energy available for particle production
still increases with increasing beam energy.

This work was supported by the Division of Nuclear Physics of the
Office of Science of the U.S. Department of Energy under contracts
DE--AC02--98--CH10886, DE--FG03--93--ER40773, DE--FG03--96--ER40981, and
DE--FG02--99--ER41121, the Danish Natural Science Research Council,
the Research Council of Norway, the Polish Ministry of Science and Information
Society Technologies(Grant no. 0383/P03/2005/29), and the Romanian
Ministry of Education and Research (5003/1999, 6077/2000). We
thank the staff of the Collider-Accelerator Division at BNL for
their excellent and dedicated work to deploy RHIC and their
support to the experiment.

\newpage
\end{document}